\documentclass[a4paper]{jpconf}
\usepackage{graphicx,aas_macros,amsmath,amssymb,color}
\usepackage[colorlinks=true,
            urlcolor=blue,
            linkcolor=blue,
            citecolor=red]{hyperref}
\begin{document}

\title{A simple and efficient solver for self-gravity in the DISPATCH astrophysical simulation framework}

\author{J P Ramsey and T Haugb\o lle and \AA\ Nordlund}

\address{Centre for Star and Planet Formation, the Niels Bohr Institute and the Natural History Museum of Denmark,
University of Copenhagen, \O ster Voldgade 5-7, DK-1350 Copenhagen, Denmark}

\ead{jramsey@nbi.ku.dk; haugboel@nbi.ku.dk; aake@nbi.ku.dk}

\begin{abstract}
We describe a simple and effective algorithm for solving Poisson's equation in the context of self-gravity within the DISPATCH astrophysical fluid framework. The algorithm leverages the fact that DISPATCH stores multiple time slices and uses asynchronous time-stepping to produce a scheme that does not require any explicit global communication or sub-cycling, only the normal, local communication between patches and the iterative solution to Poisson's equation. We demonstrate that the implementation is suitable for both collections of patches of a single resolution and for hierarchies of adaptively resolved patches. Benchmarks are presented that demonstrate the accuracy, effectiveness and efficiency of the scheme.
\end{abstract}

\section{Introduction}
\label{sec:intro}
Gravity is at the root of the formation of cosmological structure, galaxies, stars and planets. It also dominates the dynamics of many astrophysical systems, not least planets in orbit around a star or the spiral arms of a galaxy. Determining the strength of gravity at any given location requires solving for the gravitational potential due to a particular mass distribution via Poisson's equation, a second-order partial differential, elliptical equation with Dirichlet boundary conditions. In general, the solution to Poisson's equation for an arbitrary mass distribution is not analytically tractable. As such, given the importance of gravity in the universe, the accurate (and efficient) numerical solution of Poisson's equation has received considerable attention in computational astrophysics.

There exists many different approaches to numerically solve Poisson's equation. The most popular, basic methods include fast Fourier transforms (FFT), direct N-body calculations \cite{evrard1988_nbody}, multi-pole expansions \cite{couchetal2013_multipole}, relaxation methods (e.g.\ Gauss-Seidel and successive over relaxation; SOR) and multi-grid techniques \cite{press92_nr}. Many problems in astrophysics, meanwhile, have an intrinsically large dynamic spatial range (e.g.\ galaxy and star formation), which makes these problems expensive to simulate. In order to keep the cost of simulations down, considerable effort has been spent developing adaptive techniques that focus numerical resolution only when and where it is needed (e.g.\ adaptive mesh refinement; AMR; \cite{bc89}). Naturally, the combination of these adaptive techniques with the solution of Poisson's equation is desirable and indeed required in certain contexts. Examples of adaptive resolution techniques for Poisson's equation include variants or combinations of the above listed methods (e.g.\ FFT + multi-grid; \cite{collinsetal2010_enzo}) that have been modified for a specific adaptive framework or data structure (e.g.\ AMR multi-grid; \cite{trueloveetal1998,ricker2008_octtree,almgrenetal2010_castro}; tree gravity \cite{wunschetal2017_treegrav}; sink particles \cite{federrathetal2010_sinks}).

What all of these methods share in common is the need to globally synchronise information whenever a solution to Poisson's equation is undertaken. Indeed, in a large-scale, domain-decomposed simulation this usually requires substantial communication of information across domains (most commonly, MPI ranks\footnote{MPI: ``Message Passing Interface''; currently, the most common approach to distributed scientific computing.}) as the Poisson solver iterates, and a synchronisation point whereby all domains agree on the global solution. Not surprisingly, the details of how the Poisson equation is solved and how updated information is propagated between domains can have a substantial effect on the performance and scalability of a code or simulation framework.

In this contribution, we present a simple and efficient method for solving self-gravity in the DISPATCH framework. DISPATCH is a new framework for grid-based astrophysical fluid simulations \cite{dispatchI,popovasetal18,dispatchII} which incorporates novel algorithms to produce a framework with potentially unlimited parallel scaling and excellent performance even at the node level. The salient features of DISPATCH can briefly be described as follows: 1) We decompose the computational domain into small, semi-independent patches, which readily fit into cache and can be efficiently vectorised; 2) we store multiple time slices of the conservative variables, which then permits us to use individual, asynchronous time-stepping in each patch; 3) we ensure there are many more patches than hardware threads (e.g.\ OpenMP) available, thus ensuring threads keep busy; 4) work scheduling is task-based and patches most in need of updating are picked first; 5) each MPI rank only communicates with its nearest neighbours, whether they be geometric neighbours or causally-coupled ones; 6) load and communication imbalances are handled by trading entire patches to MPI neighbours on-demand and continuously. Taken together, these features prevent a very small fraction of the computational domain from dominating the cost of a simulation, they permit continuous and low-cost load balancing and they result in linear OpenMP and MPI scaling to $>10^5$ cores \cite{dispatchI}.

The modular design of DISPATCH allows for different solvers to be used depending on the set of partial differential equations being solved and the needs of the user. Currently available solvers include the RAMSES hydrodynamical (HD) and magnetohydrodynamical (MHD) solvers \cite{teyssier2002, fromangetal2006_ramses}, STAGGER solvers of 2nd, 4th and 6th order \cite{kritsuketal2011}, the ZEUS-3D solver \cite{clarke2010,rcm12} and even the PPcode solver for particle-in-cell applications \cite{haugbolleetal2013_ppcode}. In this work, we choose to apply the hydrodynamical variants of the ZEUS-3D and 2nd-order STAGGER solvers. The self-gravity algorithm presented herein is, however, independent of the choice of hydrodynamical solver.

What makes this particular algorithm novel is that it exploits the multiple time slots and asynchronous time-stepping features of DISPATCH, plus the existing infrastructure for prolongation and restriction, leading to a simple but effective means for solving Poisson's equation. Indeed, the algorithm requires no global communication or blocking synchronisation across domains.
\section{Solving Poisson's Equation}
\label{sec:poisson}
Poisson's equation for Newtonian gravity is a prototypical example of a second-order partial differential, elliptic equation constrained by boundary conditions:
\begin{equation}
  \label{eq:poisson}
  \nabla^2\Phi = 4\pi G \rho,
\end{equation}
where $\rho$ is the gas density, $G$ is the universal gravitational constant, $\Phi$ is the gravitational potential and $\nabla^2 = \nabla\cdot\nabla$ is the Laplacian operator. The term on the right hand side is commonly referred to as the ``source'' and, in the case of gravity, this is the distribution of mass.

In periodic cases, one must  offset the source term by its global average. Doing this renormalises the problem, effectively cancelling the effect of the mass in the periodic neighbours, which could otherwise lead to a divergent solution.

To wit, the form of the Poisson equation that must be solved in the periodic case is:
\begin{equation}
  \label{eq:poisson_offset}
  \nabla^2\Phi = 4\pi G (\rho - \langle\rho\rangle) = 4\pi G \rho',
\end{equation}
where $\langle\rho\rangle$ is the average value of the density over the entire domain. It is sufficient to determine this value from the initial conditions and therefore does not require a global summation. In cases with fixed boundary conditions, there is no need for such a renormalisation, and would indeed result in an incorrect solution.

As already discussed, many different methods are available to solve eq.\ \ref{eq:poisson_offset}. Direct methods (e.g.\ FFT, N-body) would be preferred because they provide an ``exact'' solution, but these approaches are often impractical or even intractable in many cases. This is particularly true when the mass is discretised onto a hierarchy of grids, e.g., when using AMR. Instead, iterative relaxation methods have proven to be the most cost-effective in this situation. We choose to solve eq.\ \ref{eq:poisson_offset} iteratively via the form:
\begin{equation}
  \label{eq:resid}
  4\pi G R = \nabla^{2}\Phi - 4\pi G\rho',
\end{equation}
where $R$ is the residual and is a direct measure of the adjustment in mass that would be required to make $\Phi$ the exact solution for a given set of boundary conditions. While the goal is to reduce the residual to zero iteratively, in practice, one instead aims to decrease its magnitude below a given tolerance level. In our case, the quantity used to measure convergence is:
\begin{equation}
  \label{eq:converge_criteria}
  \varepsilon = \frac{R}{\rho+\rho_0} , 
\end{equation}
where $\rho_0$ is a `floor' value characterising the desired tolerance in regions with very low density. When $\varepsilon < \varepsilon_0$, typically set to $10^{-4}$, convergence is signalled and the Poisson solver exits.

For solving Poisson's equation on a single DISPATCH patch, we have implemented both successive over-relaxation (SOR) with Chebyshev acceleration \cite{press92_nr} and a conjugate gradient (CG) solver \cite{ghyselsvanroose2014_pcg} with preconditioner \cite{ament2010_pcg}. One could, of course, also apply Gauss-Seidel relaxation, but since SOR is always faster there is no reason to. An important consideration for SOR is that a red-black scheme should be employed, otherwise initial symmetries that exist in a problem will not be preserved. In fact, preserving exact symmetry, in any case, may require the use of an odd number of points in non-periodic problems, and an even number of points in periodic problems.

The cost per point of the SOR and CG (with preconditioner) solvers are relatively similar for a typical number of cells per patch. For example, the average cost per cell update on an Intel Core i5 Haswell-based laptop is $\sim$0.015 $\mu$s/point/iteration. Of course, as the number of cells increases, the CG solver becomes more efficient than SOR (c.f.\ the tests of Section \ref{sub:static}).

Once we have obtained the gravitational potential, the gravitational acceleration, $\vec{g} = \nabla\Phi$, can be calculated and applied as a source term within the hydrodynamical solver. The Poisson solver is implemented independently of the hydrodynamical solver; the feedback of gravity on the gas is accomplished via accelerations. For example, when using 2nd-order staggered differences (e.g.\ ZEUS-3D and STAGGER solvers), the $x$-component of the gravitational acceleration at the cell face indexed by $(i,j,k)$ may be written as:
\begin{equation}
  \label{eq:g_staggered}
  g_{x,i} = -\partial_x \Phi = -\frac{\Phi_{i} - \Phi_{i-1}}{\Delta x} + \mathcal{O}(\Delta x^2),
\end{equation}
where $\Delta x$ is the grid spacing in the $x$-direction. Since the gravitational potential is zone-centred, this simple form gives accelerations that are face-centred and co-spatial with staggered velocity/momentum components. In the case of a centred solver, such as RAMSES, we instead apply the following centred difference formula \cite{guilletteyssier2011}:
\begin{equation}
  \label{eq:g_ramses}
  g_{x,i} = -\partial_x \Phi = -\frac{4}{3}\frac{\Phi_{i+1} - \Phi_{i-1}}{2\Delta x} + \frac{1}{3}\frac{\Phi_{i+2} - \Phi_{i-2}}{4\Delta x} + \mathcal{O}(\Delta x^4).
\end{equation}

\section{Multiple patch and multiple resolution algorithm}
\label{sec:multipatch}
When applying adaptive grid techniques to the solution of the (M)HD equations (e.g.\ \cite{bc89}), small errors are readily introduced at interfaces between grids of different resolutions. In the case of quantities which are relatively smooth, these errors can be compensated for by applying so-called ``flux corrections''. In the more extreme case that a discontinuous wave passes through a change in resolution, however, errors are unavoidably generated (see, e.g., Sect.\ 7.1 of \cite{rcm12}). However, these errors are rarely ever consequential in complex, large-scale numerical simulations and, as such, they are frequently tolerated.

Although self-gravity is clearly a global problem, the gravitational potential is typically smooth and slowly-varying slowly in time. With this in mind, in DISPATCH, we choose to take a different approach to self-gravity: We solve Poisson's equation locally, on a per patch basis, and then export the updated information to neighbouring patches. This potentially admits errors after a single call to the Poisson solver but, independent of the strategy to reduce such errors, their magnitude can always be monitored by computing the residual (eq.\ \ref{eq:resid}). If the errors become large, it is typically because the density distribution changes too much per time step, and a simple remedy is then to reduce the local time step. This effectively leads to more Poisson solutions per unit code time, and hence more frequent exchange of updated information between neighbours. Importantly, frequent exchange between neighbours propagates information about changes in the mass distribution across the domain at a rate proportional to the inverse of the (M)HD time step. Furthermore, Poisson's equation is iterated until convergence on all patches within the domain.

The first step to solving for gravity in DISPATCH is restriction and prolongation, which we term here as `downloading'. Before the Poisson solver is invoked, up-to-date information about the mass distribution and gravitational potential in neighbouring patches is downloaded. Depending on the details of the patch arrangement, this will either take the form of temporal and spatial interpolation of guard zone data from neighbours (i.e.\ prolongation) or the conservative averaging of finer resolution data from overlying child patches to the current patch (i.e.\ restriction). In the first regard, we employ linear interpolation in both time and space; higher order interpolation in space is possible, but more costly, and linear interpolation has proven more than sufficient for the current application. In the second regard, we use linear averaging. Via experimentation, we have found that using the finest resolution data available when prolonging values of $\Phi$ provides the best precision. The downloading procedure for the gravitational potential is the same as for any of the hydrodynamical variables. Indeed, that is one of advantages of our scheme: no special considerations are needed for the gravitational potential relative to the other variables.

The next step is to actually solve Poisson's equation on a patch (see above). This is accomplished after the download, but before the (M)HD equations are solved. The resulting gravitational potential is then applied as a source term in the (M)HD solver (eqs.\ \ref{eq:g_staggered}, \ref{eq:g_ramses}).

A critical part of the algorithm is to now call the Poisson solver again for the same patch but using the recently time-updated values of the density. This solution typically only takes 1 or 2 iterations to converge and the resulting gravitational potential is placed into the \emph{next} time slot. 
When using local time-stepping, a neighbour patch will generally not be at exactly the same time as the current patch and will therefore require some interpolation or extrapolation in time to fill its ghost cells before it can take a step forward in time; by providing a forward \emph{prediction} for $\Phi$, the neighbour can fill its ghost cells and proceed with its update.

Since we have access to multiple time slices of $\Phi$ for each patch in DISPATCH (see \cite{dispatchI}), it is possible to make a (e.g.\ linear, quadratic, or higher order) extrapolation in time for $\Phi$ from these time slices that can then serve as an initial guess for the next invocation of the Poisson solver on a given patch. Experimentation shows that a linear extrapolation in time for a single, isolated patch can reduce the number of iterations in a subsequent Poisson solve by up to a factor of three. Quadratic temporal extrapolation, meanwhile, typically reduces the iteration count by a further $\sim$10\%. This feature is most effective when the potential changes significantly between time steps (e.g.\ a quickly moving massive clump). In certain pathological cases, this feature results in non-convergence of the Poisson solver, however, and thus it is implemented as an optional feature.

Finally, in our algorithm, if an initial guess for $\Phi$ is not available, a short ``burn-in'' phase may be desirable to obtain a globally consistent $\Phi$ before the dynamics are activated. This is accomplished by running DISPATCH with the dynamics deactivated and only solving for the gravitational potential (e.g.\ Section \ref{sub:static}). The procedure followed is the same as when dynamics are activated, and so no special modifications are necessary for this phase. Furthermore, it remains a local process; no explicit global communication of information is required.

\section{Verification}
\label{sec:validate}
\subsection{Static gravitational potential}
\label{sub:static}
As a first test of our self-gravity algorithm, we investigate a slightly modified version of the static, analytical gravitational potential of \cite{guilletteyssier2011}. This test employs a 2D, analytic, static and purely radial gravitational potential given by:
\begin{equation}
  \label{eq:phi_exact}
  \Phi_\mathrm{e}(r) = \ln \left[ \left(\frac{r}{r_0}\right)^{2} + 1\right] + 1,
\end{equation}
where $r^2 = x^2 + y^2$ is the radial coordinate and $r_0$ is the parameter that determines the radial ``concentration'' of the potential. We have slightly modified this expression relative to \cite{guilletteyssier2011} so that, at the centre of the domain ($r = 0$), it has a non-zero value. This has no effect on the resulting density distribution or the quality of the solution, but simplifies the comparison of numerical and analytical solutions near $r = 0$.

The resulting density distribution, obtained by applying Equation (\ref{eq:poisson}) to $\Phi_\mathrm{e}$, is:
\begin{equation}
  \label{eq:d_exact}
  \rho_\mathrm{e}(r) = \frac{4r_0^{2}}{\left(r^{2} + r_{0}^{2}\right)^{2}}.
\end{equation}

The test is set up by initialising the density to eq.\ (\ref{eq:d_exact}) everywhere and setting the gravitational potential to eq.\ (\ref{eq:phi_exact}) in the boundaries of the computational domain, but zero elsewhere. We set the domain size to $(x,y) \in [-0.5,0.5]$ and the concentration parameter to $r_0 = 0.1$. We then proceed to solve for $\Phi$ iteratively on every patch. A convergence tolerance of $10^{-4}$ was employed for the Poisson solver. As described above, in between calls to the Poisson solver, updated boundary information is retrieved from neighbouring patches. We alternate between Poisson solves and downloading until the global value of the $L_\infty(\Phi)$-norm converges. As this is a time-independent test, time extrapolation in the gravitational potential was not used.

Furthermore, because this test does not require hydrodynamics or even a time step, we have exploited the modular and hierarchical structure of DISPATCH to implement a shell `solver' for Poisson's equation that leverages the framework without needing to include any code for a hydrodynamical solver.

We consider three configurations: First, a single patch with varying resolution, from $16^2$ to $2048^2$ cells. Second, the domain is split into either $3^2$ or $9^2$ patches of equal resolution and tests run for $4^2$ to $256^2$ cells per patch. Third, a set of nested, refined grids with a maximum refinement level\footnote{The ``root'' grid, which encompasses the entire domain, has level = 1.} of 3 or 4 but a constant number of cells per patch. Each nested level is comprised of $3^2$ patches centred on the origin and refined relative to the previous level by a factor of three. Tests were run with $5^2$ up to $129^2$ cells per patch. In the nested patches set up, a maximum refinement level of 3 (4) is equivalent to covering the domain with $9^2$ ($27^2$) patches.

\begin{figure}[ht]
  \includegraphics[width=0.38\linewidth]{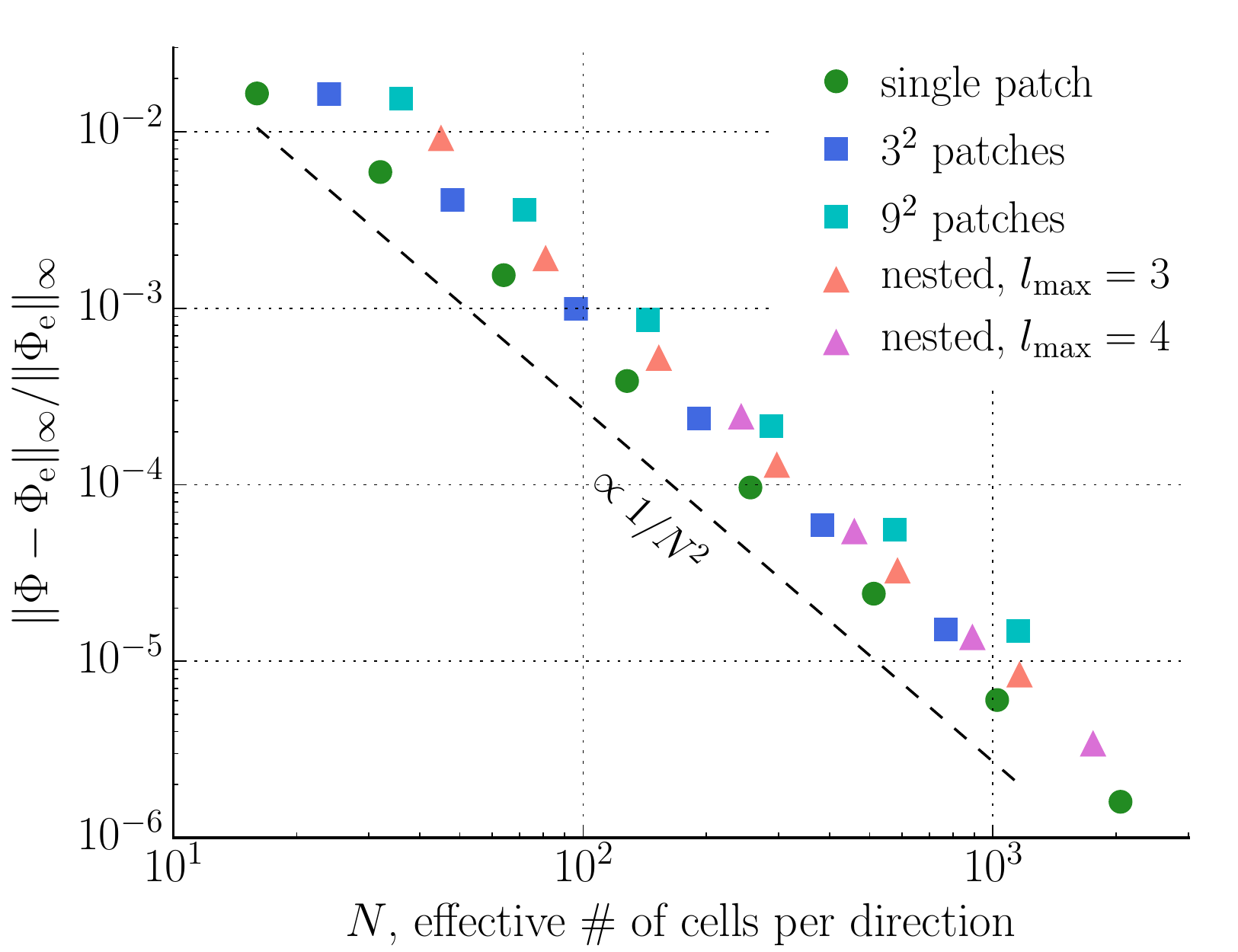}
  \includegraphics[width=0.38\linewidth]{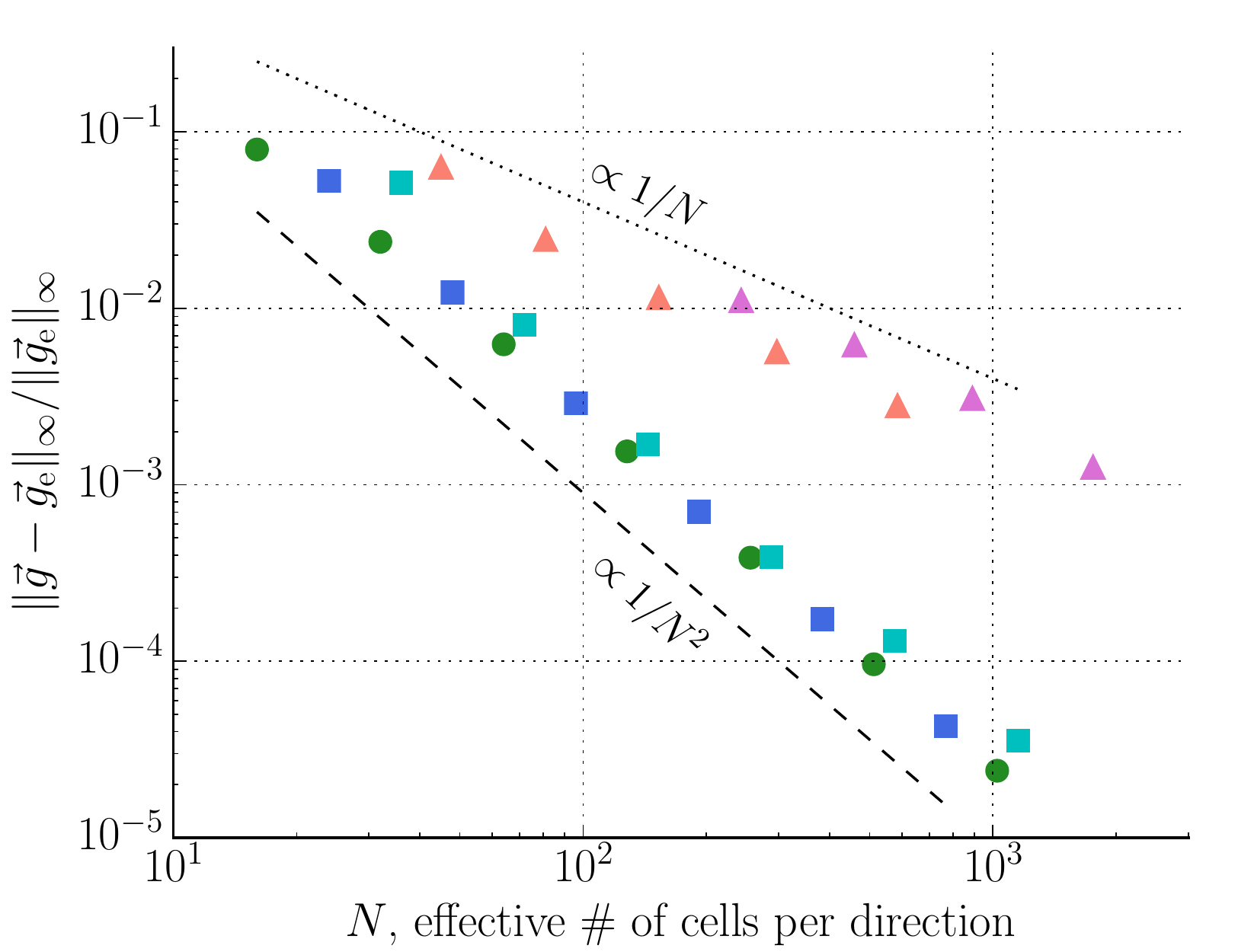}
  \includegraphics[width=0.22\linewidth]{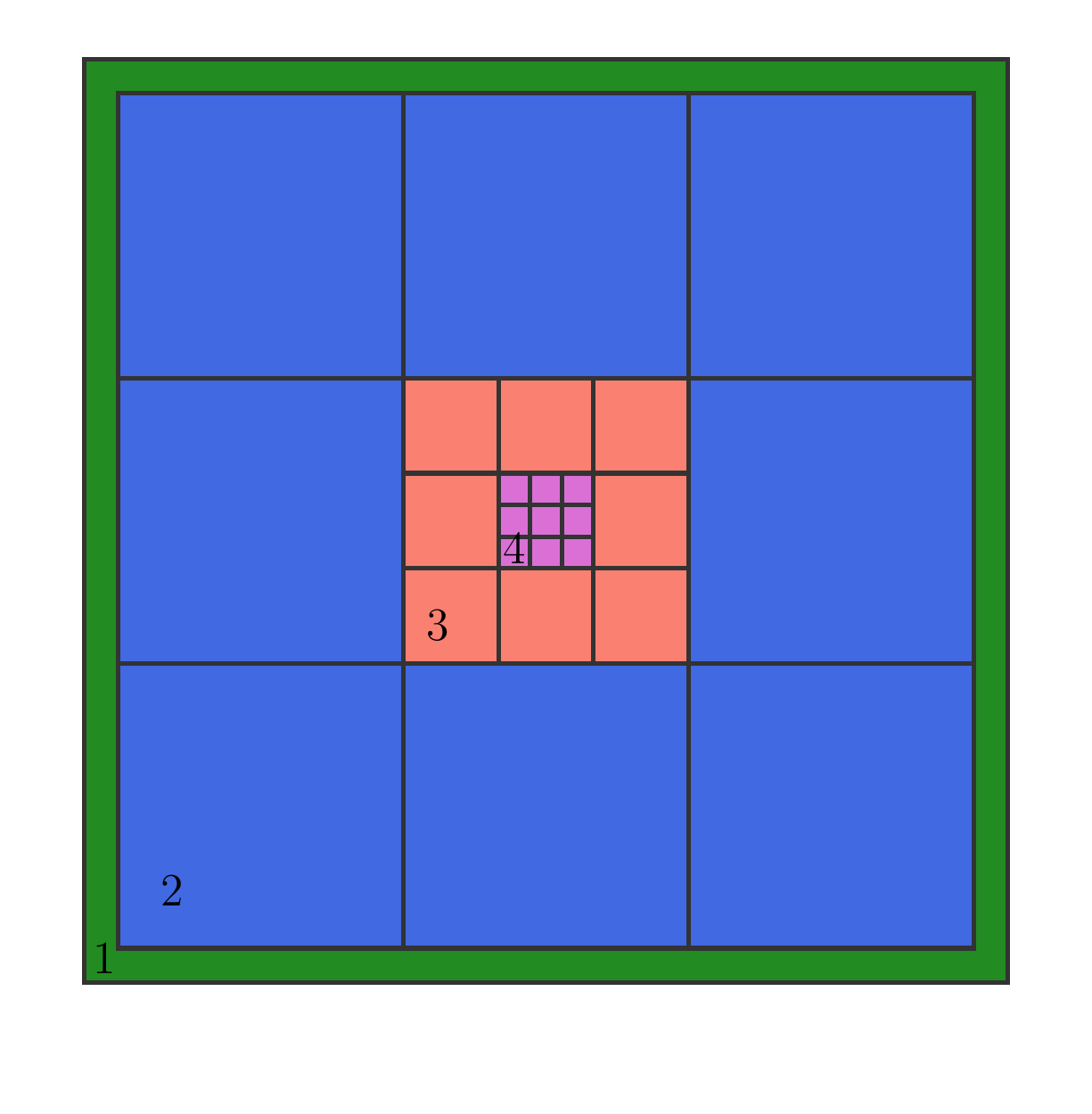}
  \caption{\emph{Left:} The $L_{\infty}$-norm in the gravitational potential for the static potential problem of Sect.\ \ref{sub:static} as a function of (effective) resolution. The straight line denotes an $L_{\infty}$-norm which is proportional to $1/N^2$, where $N$ is the effective number of cells. \emph{Middle:} The $L_{\infty}$-norm in the gravitational acceleration given by eq.\ \ref{eq:g_staggered}, as a function of resolution. \emph{Right:} A schematic representation of the nested patch hierarchy used for $l_\mathrm{max} = 4$. Numbers denote level; black lines denote patch boundaries.}
  \label{fig:staticpot}
\end{figure}

Figure \ref{fig:staticpot} shows the resulting $L_\infty$-norm values in both the gravitational potential and acceleration for the three scenarios at different effective resolutions: a single patch of uniform resolution (circles), a domain sub-divided into multiple patches (squares), and nested sets of patches of increasing resolution (triangles). With respect to the gravitational potential, all three scenarios show that the $L_\infty$-norm decreases proportional to the number of cells squared, demonstrating that our algorithm converges at second-order, and that the $L_\infty$-norm does not vary significantly between configurations. As for the gravitational acceleration, $\vec{g}$ our algorithm retains second-order convergence when all patches are of the same resolution. However, once patches of different resolution are introduced, we observe that the convergence in the gravitational acceleration degrades to first-order. As noted in other works (e.g.\ \cite{guilletteyssier2011}), this is a result of using linear interpolation for filling guard zones at changes in resolution, and adopting a higher-order interpolation in $\Phi$ would improve the convergence rate.

\subsection{The Truelove experiments}
\label{sub:truelove}
The collapse of a gaseous clump under its own gravity is critical to understanding star, disk and planet formation and its dynamics has therefore been well-studied under a wide range of conditions (e.g.\ \cite{hoyle1953, hunter1962,nordlund2014,burkertbodenheimer1993,bateburkert1997,federrathetal2010_sinks}).

The criterion for runaway gravitational collapse is originally due to Jeans \cite{jeans1902}, who found that density perturbations on scales larger than:
\begin{equation}
  \label{eq:jeanslength}
  \lambda_\mathrm{J} = \sqrt{\frac{\pi c_\mathrm{s}^2}{G\rho}}
\end{equation}
are unstable. With respect to numerical simulations of gravitational collapse, \cite{trueloveetal1997, trueloveetal1998} further determined that, in order to avoid numerical fragmentation, one must ensure that the so-called Jeans number, 
\begin{equation}
  \label{eq:jeansnumber}
  J = \frac{\Delta x}{\lambda_\mathrm{J}},
\end{equation}
where $\Delta x$ is the cell size, is kept below a value of $J_\mathrm{max}$ = 1/4.

To test the combination of self-gravity and hydrodynamics, we follow \cite{trueloveetal1998} and place a massive clump of radius $R_0$ that is, by definition, gravitationally unstable, at the centre of a 3D Cartesian domain. The density of the clump is given by $\rho_0 (1 + A\cos(m\vartheta))$ for $r \leq R_0$ and by $\rho_0 / \chi_\rho$, otherwise, where $\rho_0$ is the uniform density of the clump, $A$ is the amplitude of a possible perturbation, $m$ is the mode of the perturbation, $\vartheta$ is the polar angle in the $x-y$ plane and $\chi_\rho = 100$ is the density contrast between clump and ambient medium. In addition to the density, these clumps are characterised by their mass, $M = 1M_\odot$, and their thermal and rotational energies with respect to gravity:
\begin{align}
  \label{eq:truelove_alpha}
  \alpha = \frac{E_\mathrm{therm}}{E_\mathrm{grav}} & = \frac{5}{2}\left(\frac{3}{4\pi \rho_0 M^2}\right)^{1/3}\frac{c_\mathrm{iso}^2}{G};\\
  \label{eq:truelove_beta}
  \beta = \frac{E_\mathrm{rot}}{E_\mathrm{grav}} & = \frac{1}{4\pi}\frac{\Omega^2}{G\rho_0}.
\end{align}

For a given initial clump density, the gravitational free-fall time,
\begin{equation}
  \label{eq:tfreefall}
  t_\mathrm{ff} = \sqrt{\frac{3\pi}{32G\rho_0}},
\end{equation}
can be used to estimate the time scale of collapse.

The criteria for dynamic refinement applied here is that the Jeans number (eq.\ \ref{eq:jeansnumber}) must remain below 1/4 in every cell. If this criteria is violated, a patch is split into $3^3$ child patches, each with the same number of cells as the parent, and the parent data is prolonged onto the new patches.

We attempted three different configurations drawn directly from \cite{trueloveetal1998}: i) a non-rotating clump of initially uniform density; ii) a uniformly rotating clump of uniform density; iii) a slightly perturbed (10\% amplitude in the $m = 2$ mode) but uniformly rotating clump. The parameters employed can be found in Table \ref{tab:truelove}.

We employed both the ZEUS-3D solver and the 2nd-order STAGGER solver. In the current configuration, both employ the internal energy equation and use a Courant factor of 0.2. The solver-specific artificial viscosity parameters employed for ZEUS-3D are \verb+qlin+ = 0.1 and \verb+qcon+ = 1 \cite{clarke2010}, while the relevant hyper-viscosity parameters for STAGGER are $\nu_1$ = 0.2 (viscous stress tensor coefficient), $\nu_2$ = 1.0 (artificial pressure coefficient) \cite{kritsuketal2011}. We furthermore used $17^3$ cells per patch, a refinement ratio of three and a convergence tolerance of $10^{-4}$ for the Poisson solver.

\begin{figure}[ht]
  \centering
  \includegraphics[width=0.49\linewidth]{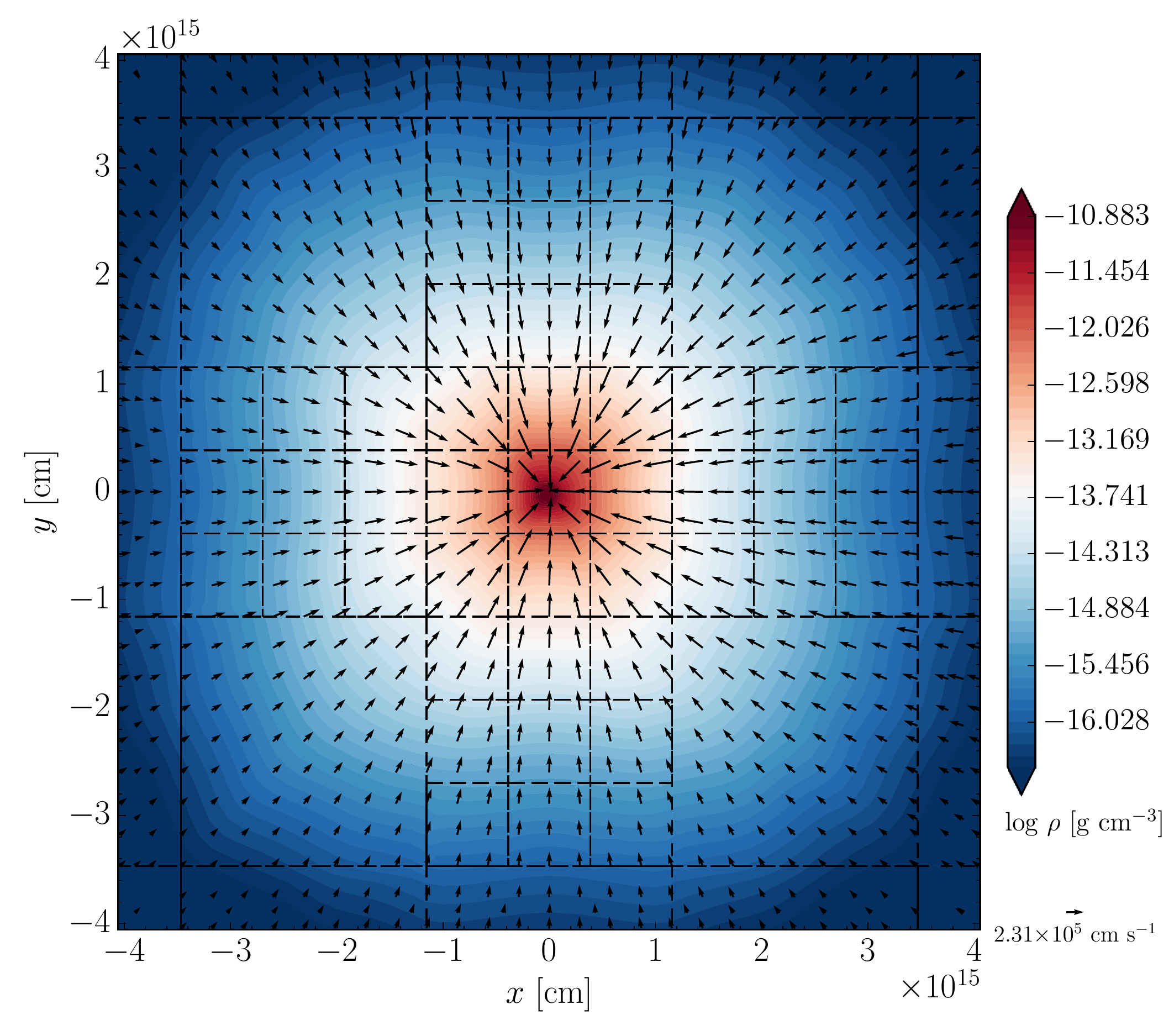}\includegraphics[width=0.49\linewidth]{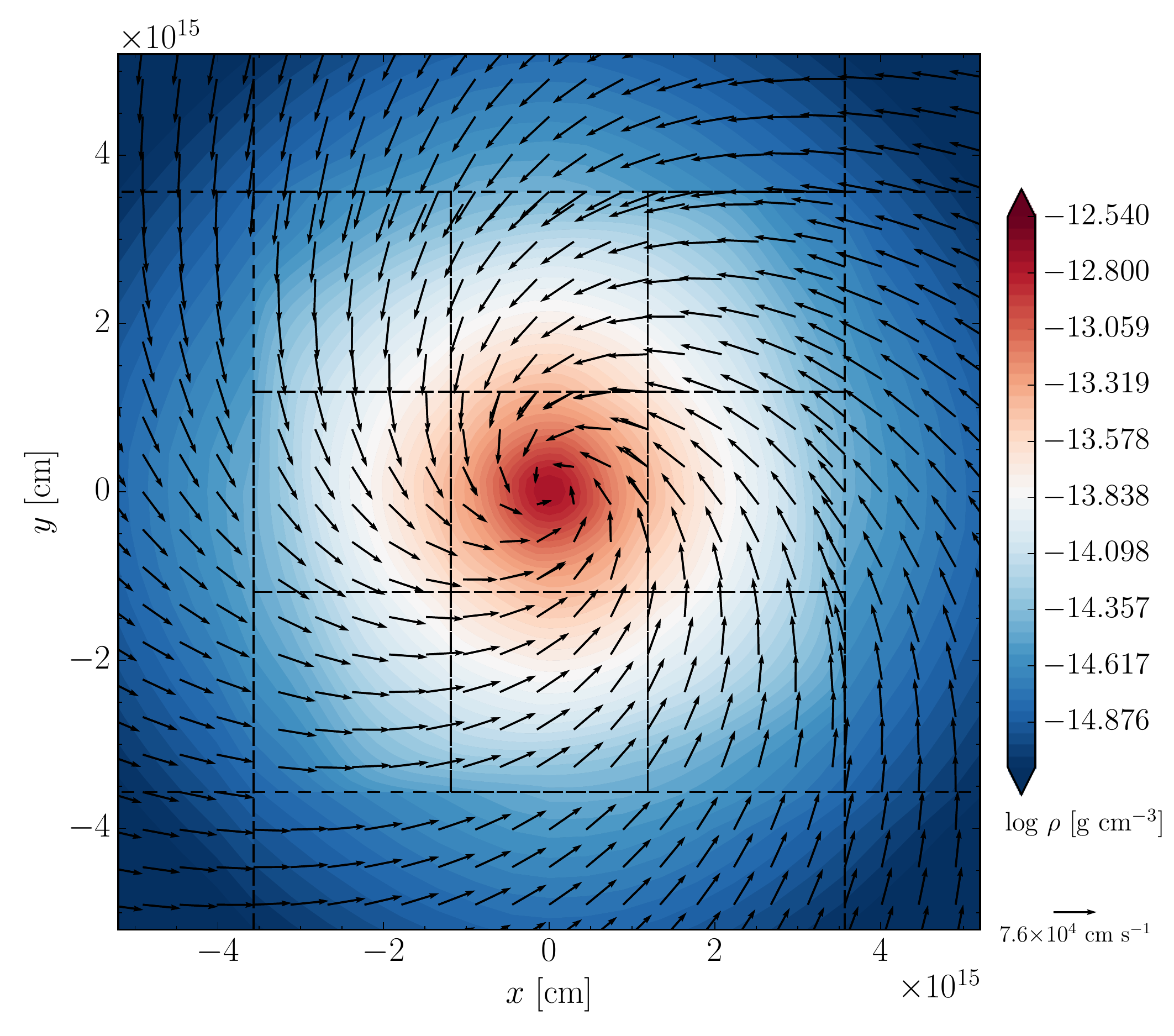}
  \includegraphics[width=0.49\linewidth]{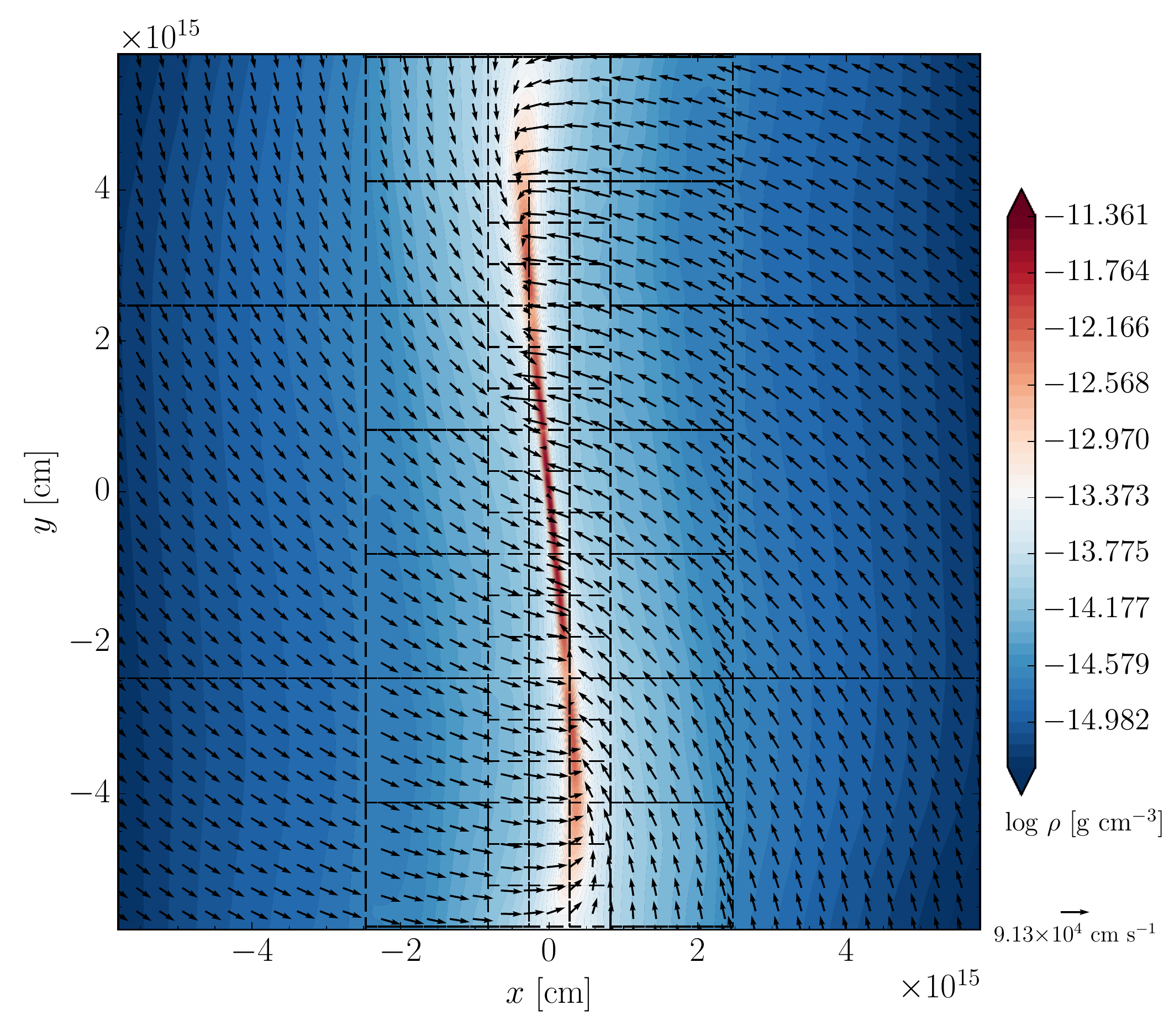}\includegraphics[width=0.49\linewidth]{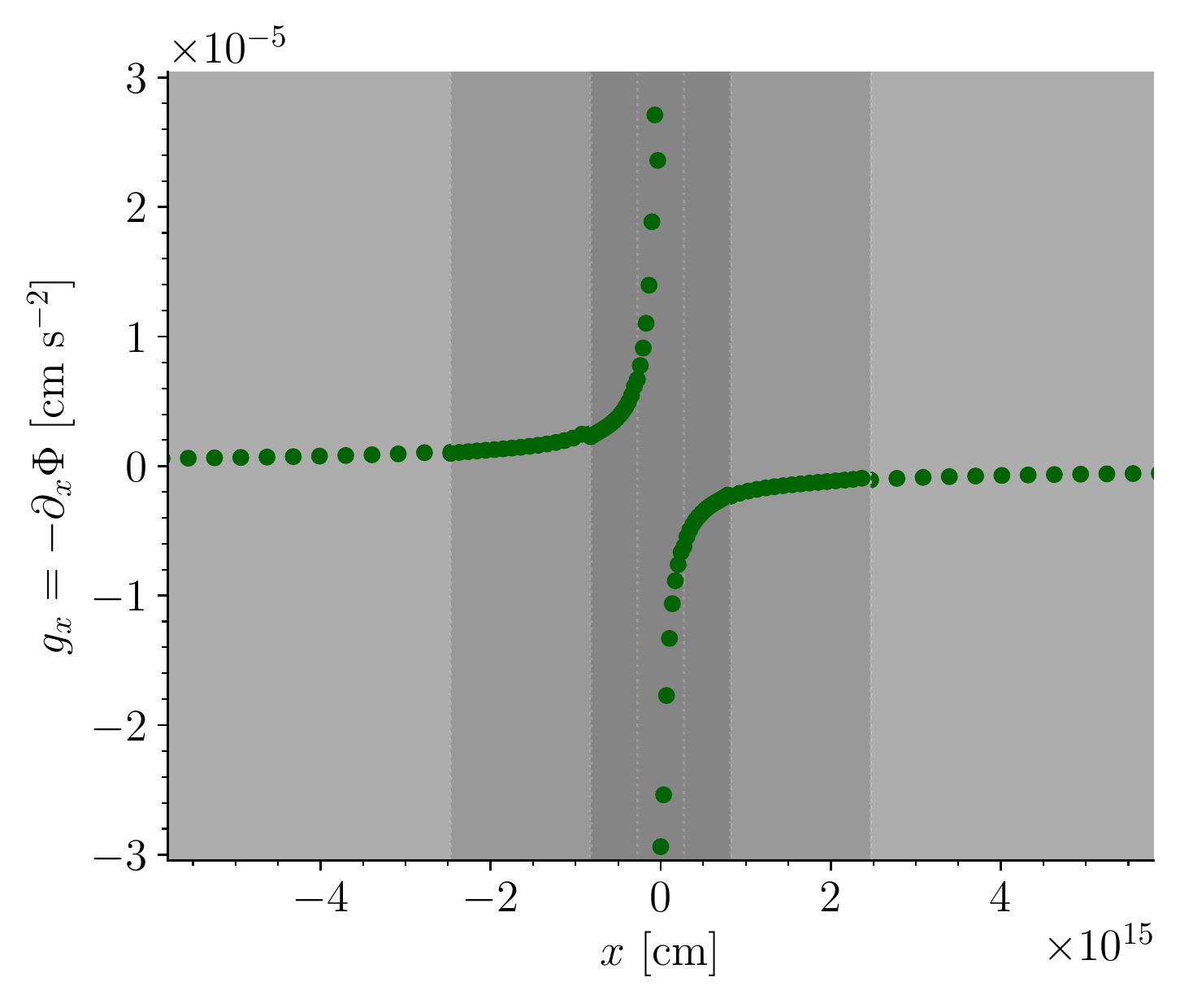}
  \caption{Truelove collapse tests. Plotted is an $x-y$ slice of the logarithm of density at $z = 0.0$. Black arrows denote velocity vectors. Dashed lines indicate patch boundaries. \textit{Top left:} Uniform, non-rotating collapse at $t = 1.036t_\mathrm{ff}$. \textit{Top right:} Uniform, rotating collapse at $t = 1.239t_\mathrm{ff}$. \textit{Bottom left:} Perturbed ($m = 2$, 10\% amplitude), rotating collapse at $t = 1.321t_\mathrm{ff}$. \textit{Bottom right:} The $x$-component of the gravitational acceleration (eq.\ \ref{eq:g_staggered}) through $y = z = 0.0$ in the perturbed Truelove experiment. Successively darker shading indicates levels 4--6 while vertical dashed lines denote patch boundaries.}
  \label{fig:truelove}
\end{figure}

As can be seen in Figure \ref{fig:truelove}, our benchmark results compare favourably with those of \cite{trueloveetal1998}. There are differences in the rate of collapse, but the density and velocity structures are very similar. In the uniform, non-rotating case, by $t = 1.036t_\mathrm{ff}$, the density has increased from its initial value by 4.3 orders of magnitude to $\rho_\mathrm{max} = 10^{-10.7}$ g cm$^{-3}$ and we reach a maximum effective resolution of 459 cells ($l_\mathrm{max} = 4$). Similarly, for the uniform, rotating case, the density increases by 5.2 orders of magnitude to $\rho_\mathrm{max} = 10^{-12.6}$ g cm$^{-3}$ after $1.239t_\mathrm{ff}$ and we reach an effective resolution of 1377 cells ($l_\mathrm{max} = 5$). Finally, in the 10\% perturbation + rotating test, by $1.321t_\mathrm{ff}$, the maximum density has reached $\rho_\mathrm{max} = 10^{-11.4}$ g cm$^{-3}$, an increase of 6.04 orders of magnitude, and we attain an effective resolution of 4131 cells ($l_\mathrm{max} = 6$). The bottom right panel of Figure \ref{fig:truelove} shows a slice of the $x$-component of the gravitational acceleration through the center ($y = z = 0$) of the perturbed Truelove experiment. The profile appears smooth across patch boundaries, indicating that our algorithm is approximating the gravitational force reasonably. However, we do observe changes in the slope of the acceleration (at the $<$10\% level) at jumps in resolution. Therefore, we are currently investigating the use of higher-order interpolations for $\Phi$ in the boundaries.

Differences in our results with respect to \cite{trueloveetal1998} can be attributed to two factors: i) We are employing fixed rather than periodic boundary conditions and ii) we are using a box size of $8R_0$ versus $4R_0$. The absence of `mirror' images due to periodic boundaries, combined with a larger box size, implies that our set up should collapse somewhat faster than the simulations presented in \cite{trueloveetal1998}. Indeed, with the exception of the uniform, rotating collapse, our simulations attain higher densities than at the times stated in \cite{trueloveetal1998}.

\begin{table}
\centering
\begin{tabular}{rccc}
\hline
 & Non-rotating & Rotating & \begin{tabular}{@{}c@{}}Rotating\\ + perturbed\end{tabular} \\
\hline\\[-2ex]
$\rho_0~ [\mathrm{g\, cm^{-3}}]$ & $10^{-15}$ & $10^{-17.9}$ & $10^{-17.4}$ \\
$\alpha$ & 0.0475 & 0.54 & 0.26 \\
$\beta$ & 0 & 0.08 & 0.16 \\
$\chi_p = \tfrac{P_\mathrm{clump}}{P_\mathrm{ambient}} $ & 1 & 1 & 10 \\[0.6ex]
\hline\\[-2.0ex]
$R_0\, [\mathrm{cm}]$ & $7.80\times 10^{15}$ & $7.23\times 10^{16}$ & $5.0\times 10^{16}$ \\
$t_\mathrm{ff}$ [s] & $6.65\times 10^{10}$ & $1.87\times 10^{12}$ & $1.05\times 10^{12}$ \\
\hline
\end{tabular}
\caption{Parameters and derived values used in the Truelove tests. In all cases, we employ a box size of $8R_0$, $\gamma = 1.0001$, $\chi_\rho = \rho_\mathrm{clump} / \rho_\mathrm{ambient} = 100$ and a cloud mass of $1M_\odot$. See the text for additional details.}
\label{tab:truelove}
\end{table}
\section{Summary}
\label{sec:summary}
Herein, we have presented a method for solving Poisson's equation on adaptive resolution grids for self-gravity that does not require any explicit global communication. The approach is built around the concept of multiple individual patches that only exchange information via prolongation and restriction operations; each patch solves gravity locally and then exchanges updated information with its neighbours. The approach exploits the local time-stepping and time slice storage features of DISPATCH to provide time interpolated and extrapolated values of the gravitational potential. There is also the option to increase efficiency by using forward time prediction and by using OpenMP to solve Poisson's equation simultaneously on more than one patch.

Using the static potential benchmarks of \cite{guilletteyssier2011} and the gravitational collapse benchmarks of \cite{trueloveetal1998}, we have demonstrated that Poisson's equation can be efficiently solved on a collection of semi-independent patches, and that a global communication or synchronisation step is not required to obtain an accurate solution.

In the Truelove collapse experiments, the DISPATCH self-gravity algorithm accounts for roughly 20\% of the total cost during a simulation. Slightly less than half of the remaining cost is due to the hydrodynamical solver, while $\sim$25-30\% is due to downloading. A portion of the downloading cost originates with the self-gravity algorithm because we make an intermediate download after the (M)HD step, but before the predictive solution of Poisson's equation; in experiments without self-gravity, the relative cost of downloading is reduced to $\sim$15\%.

The first scientific application of this self-gravity algorithm is towards models of planet formation in protoplanetary disks. More specifically, we are currently investigating the self-gravity of combined gas+dust fluids and their potential for gravitational instability. We are also currently implementing additional experiments to test this algorithm under a wider range of conditions including, for example, an equal-mass binary with a moving centre-of-mass. The results of these benchmarks will be featured in future work. Finally, DISPATCH, and the self-gravity algorithm presented here, will be made open-source and publicly available in the near future.

\ack
The work of TH was supported by the Sapere Aude program of the Danish Council for Independent Research (DFF). The work of \AA N was supported by grant 1323-00199B from the DFF. The Centre for Star and Planet Formation is funded by the Danish National Research Foundation (DNRF97). Storage and computing resources at the University of Copenhagen HPC centre, funded in part by Villum Fonden (VKR023406), were used to carry out some of the simulations presented here.
\section*{References}
\bibliographystyle{iopart-num}
\bibliography{references}

\end{document}